\documentstyle[preprint,prb,aps]{revtex}
\begin{document}
\draft
\title{Modulated structures stabilised by spin softening: an expansion in
inverse spin anisotropy}
\author{F. Seno$\null^1$ and J.M. Yeomans$\null^2$}
\address{(1) INFM-Dipartimento di Fisica "G. Galilei", Via Marzolo 8, I-35131,
Padova, Italy.\\
(2) Department of Physics, University of Oxford,
1 Keble Road, Oxford, OX1 3NP, England.\\}
\date{\today}
\maketitle

\begin{abstract}
We develop an analytic approach which allows us to study
the behaviour of spin models with competing interactions
and $p$-fold
spin anisotropy, $D$, in the limit where the pinning potential which
results from $D$ is large.
This is an expansion in inverse spin anisotropy which must be carried
out to all orders where necessary.
Interesting behaviour occurs near where the boundary
between different ground states is infinitely degenerate for infinite $D$.
Here as $D$ decreases and the spins are allowed to soften, we
are able to demonstrate
the existence of several different behaviours ranging from
a single first-order boundary to infinite
series of commensurate phases. The method is illustrated by considering
the soft chiral clock model and the
soft clock model
with first- and second-neighbour competing interactions. In the latter
case the results are strongly dependent on the value of $p$.
\end{abstract}
\newpage

\section{Introduction}

There are many examples of long-period phases in nature. These include
the ferrimagnetic phases of the rare earths, long-period atomic
ordering in binary alloys and polytypism, the possibility of many
different forms of long wavelength structural order in some minerals.

The underlying physical mechanism responsible for the formation of
long-period structures is often the existence of competing interactions.
Therefore it is of interest to understand the properties of model
systems with such competition. Perhaps the simplest of these
is the ANNNI model, a ferromagnetic Ising model with
second-neighbour antiferromagnetic interactions along one lattice
axis. Despite its simplicity the ANNNI model has a very rich phase
structure with infinite sequences of commensurate and incommensurate
phases at finite temperatures\cite{Selke92,Yeo88}.

If continuous spins are considered similar structures can occur even
in the ground state. Banerjea and Taylor\cite{BT84}
performed numerical work on the chiral X-Y
model with $p=2$-fold spin anisotropy $D$ and showed that long-period
phases are stable. Chou and Griffiths\cite{CG86}
later proved that an infinite number of
commensurate phases appear as ground states for $p\geq 3$.
Numerical work has also
shown that the X-Y model with first- and second-neighbour competing
interactions and $p=6$-fold spin anisotropy has a highly complicated
ground-state phase diagram\cite{SYHK94}.

Note that a common feature of these models is the spin anisotropy $D$.
As $D$ is increased from zero to infinity the spins are confined to
increasingly deep potential wells and the model crosses over from a
continuous to a discrete spin limit. For example the X-Y model with
$p$-fold spin anisotropy becomes a $p$-state clock model for
$D=\infty$.

For infinite spin anisotropy
the ground state typically comprises a few short-period
phases. The boundaries between the phases can
either correspond to a first-order
transition where only the neighbouring phases are stable or a
multiphase point at which an infinite number of phases are
degenerate\cite{FS80}. As $D$ decreases from infinity the states at the
multiphase point can either remain degenerate with the point becoming
a multiphase line or their energy can be differentiated as the spins
soften. In the latter case some or all of the degenerate phases
may become ground
states in their own right and typically
a fan of phases springs from the multiphase point. Indeed we shall
see that in many respects $D^{-1}$ behaves in a way akin to temperature
with the spin softening playing the part of entropic fluctuations.

The large $D$ region of the phase diagram is difficult to explore
numerically because the phase sequences can be very complicated and
the widths of the stable phases small. Therefore
our aim in this paper is to describe an analytic technique which is
useful in helping to understand the ground state of models with
competing interactions for large spin anisotropy $D$.
This is an expansion in $1/D$ taken to all orders where necessary.
A short paper has summarised some of the results of the
calculations\cite{SY94}. Here our aim is to describe the technical details
of the expansion.

In Section \ref{2.0} of the
paper we explain the approach in some detail for the chiral X-Y model with
$p$-fold spin anisotropy. In \ref{2.1} the model is defined and our
notation introduced. The energy differences which are central to the
argument are defined in \ref{2.2} and their dependence on the deviation of
the spins from their positions at $D=\infty$ is calculated.
\ref{2.3} descibes the convenient labelling of the spin states which allows a
calculation of the energy differences to leading order presented in
\ref{2.4}. From
these we are able to show, in agreement with Chou and Griffiths\cite{CG86}
that all
possible phases are stable near the multiphase point for $p\ge 3$.
In Section \ref{2.5} the widths of the long period phases are calculated.

The calculation is repeated in Section \ref{3.0} for the X-Y model with first-
and second-neighbour competing interactions and $p$-fold spin
anisotropy. This is an involved calculation because of the existence
of the second-neighbour interactions.
Previous numerical results for this model for $p=6$ were
unable to proble the large $D$ limit\cite{SYHK94}.
We find that the behaviour near the multiphase points is complicated
and highly dependent on $p$.

The results are summarised and discussed in Section \ref{4.0}.

\section{The Chiral XY Model}
\label{2.0}
\subsection{Definitions and notation}
\label{2.1}
The approach is most easily explained by considering its application
to the chiral XY model with $p$-fold spin anisotropy.  This is
described by the Hamiltonian
\begin{equation}
{\cal{H}} = \sum_i \{-J ~\cos (\theta_{i-1} - \theta_i + \Delta) -
D(\cos ~p \theta_i-1)/p^2 \}
\label{a}
\end{equation}
where the $\theta_i$ are angular variables which can take values
between 0 and 2$\pi$ lying on the sites $i$ of a one-dimensional lattice.

Note that the Hamiltonian (\ref{a}) is invariant under the
transformation
\begin{equation}
\Delta \rightarrow \Delta' = \Delta + 2 \pi m/p,
\end{equation}
for any integer $m$ given the reidentification
\begin{equation}
\theta_i \Rightarrow \theta_i' = (\theta_i + 2 \pi m i /p)
\end{equation}
and therefore we may restrict our attention to $0 \leq \Delta < 2
\pi/p$.  Moreover the system is invariant under
\begin{eqnarray}
\Delta  \Rightarrow  \Delta' &=& 2 \pi/p - \Delta, \nonumber\\
\theta_j  \Rightarrow \theta_j' &=& (-\theta_j + 2 \pi j/p).
\label{b}
\end{eqnarray}
Thus the phase boundaries for $\Delta > \pi/p$ are related to those for
$\Delta < \pi/p$ by reflection in the line $\Delta = \pi/p$.  However, the
phases themselves must be identified differently within the two
regimes according to (\ref{b}).

For $D = \infty$ the spins are restricted to discrete values $2 \pi
n_i/p$ where $n_i = 0, 1, ~ \ldots ~p-1$ and the Hamiltonian (\ref{a})
becomes that of the $p$-state chiral clock model.  The ground state is
well known in this limit.  For $0 \leq \Delta \leq \pi/p$ it is
ferromagnetic whereas for $\pi/p  \leq \Delta \leq 2 \pi/p$, $n_{i+1}
= n_{i}+1$, the increase in the chirality $\Delta$ favouring a twist
in the spin ordering.  At $\Delta = \pi/p$ itself the ground state is
infinitely degenerate, with any phase for which $n_{i+1} - n_i = 0$
or 1 for all $i$ having equal energy.  Such a point is often termed
a multiphase point.  It is expected on the basis of previous work
that as $D$ decreases from infinity the degeneracy will be lifted.
Our aim is to explore the phase structure through an expansion in
$D^{-1}$. We consider $p \geq 3$.

To this end we require a notation capable of distinguishing the
different phases stable at the multiphase point.  Typically a
stable ground state will consist of a sequence of bands where
$n_{i-1} - n_i = 0$ separated by walls with $n_{i-1} - n_i = 1$.
[ $\ell_1 \ell_2 \ldots \ell_m$] will be used to describe a phase where the
repeating sequence consists of $m$ bands of length $\ell_1, \ell_2 \ldots
\ell_m$.  It may be helpful to list some examples for $p=6$
\begin{eqnarray}
{[}12{]} &&\;\;\; \ldots \mid 0 \mid 11 \mid  2 \mid  33 \mid \ldots
\nonumber \\
{[}23^2{]}&& \;\;\;\ldots \mid 00  \mid 111 \mid 222  \mid 33 \mid 444
\mid 555 \mid\ldots \nonumber \\
{[}1{]} && \;\;\;\ldots\mid 0 \mid 1 \mid  2 \mid 3 \mid 4 \mid 5\mid
\ldots \nonumber \\
{[}\infty{]}&& \;\;\;\ldots 0 00000 0 \ldots
\label{eqn5}
\end{eqnarray}
where a vertical line is used to denote a wall.  In the subsequent
text we shall use the term $l$-band to describe a band of length $l$
spins.  For example [23$^2$] consists of a 2-band followed by two 3-bands.

\subsection{The Energy Differences}
\label{2.2}
The goal is to establish which of the infinite number of phases
degenerate at the multiphase point remain stable for finite $D$.
This is done by using an expansion in inverse spin anisotropy,
$D^{-1}$.  The difficulty is that in order to check the stability
of all commensurate phases certain terms must be calculated at all
orders in $D^{-1}$.

However, the relevant terms can be identified and the phase diagram
constructed inductively using an argument first developed by Fisher
and Selke to study the phase diagram of the axial nearest neighbour
Ising model using a low temperature expansion\cite{FS80}.  We summarise their
argument here.

Consider two phases [$\alpha$] and [$\beta$] which share a common
boundary at a given order of a series expansion.  Fisher and Selke
showed that the first phase which can appear between them as the
expansion is taken to higher orders is [$\gamma$]=[$\alpha \beta$].
To check
whether this phase does indeed appear the important energy difference
is
\begin{equation}
\Delta E \equiv n_{[\gamma]} ~E_{[\gamma]} ~-~ n_{[\alpha]}~
E_{[\alpha]} ~-~ n_{[\beta]}~ E_{[\beta]}
\label{eqn6}
\end{equation}
where $E_{[\alpha]}, E_{[\beta]}, E_{[\gamma]}$ are the ground
state energies per spin and $n_{[\alpha]}, n_{[\beta]}, n_{[\gamma]}$
the number of spins per period of $[\alpha], [\beta]$ and $[\gamma]$
respectively.

There are three possibilities:\\
(i)$\;\;\Delta E > 0$ and the boundary between $[\alpha]$ and $[\beta]$
remains stable to all orders.\\
(ii)$\;\;\Delta E < 0$ and $[\alpha \beta]$ appears as a stable phase in
the vicinity of the $[\alpha]: [\beta]$ boundary. The analysis
must recommence about the new $[\alpha]:[\alpha\beta]$ and
$[\alpha \beta]: [\beta]$ boundaries.\\
(iii)$\;\;\Delta E = 0$ and  $[\gamma]$ remains degenerate on the
$[\alpha] : [\beta]$ boundary to all orders.

To explore the phase structure our goal is to calculate
$\Delta E$ and investigate its sign.  To this end we write
\begin{equation}
\theta_i = \theta^0_i + \tilde{\theta}_i
\label{eqn7}
\end{equation}
where $\theta^0_i$ is the value of the spin $\theta_i$ for
$D = \infty$ and expand the Hamiltonian (\ref{a}) to quadratic order
in the $\{\tilde{\theta}_i\}$.
\begin{equation}
\tilde{{\cal{H}}} = {\cal H} \mid_{D = \infty} + \sum_i
\{  J c_i^\theta (\tilde{\theta}_{i-1} - \tilde{\theta}_i +
s_i^\theta/c_i^\theta)^2 /2 +  D \tilde{\theta}^2_i/2 -
J (s_i^\theta)^2/(2 c_i^\theta) \} \label{c}
\end{equation}
where
\begin{eqnarray}
c_i^\theta = \cos (\theta^0_{i-1} - \theta^0_i + \Delta),&&
\;\;\;s_i^\theta = \sin (\theta^0_{i-1} - \theta^0_i + \Delta).
\label{eqn9}
\end{eqnarray}
We shall henceforth work with the quadratic approximation~(\ref{c}) to the
chiral XY model.  To leading order this gives the same
results for the energy differences as the full Hamiltonian (\ref{a}).

In equilibrium the energy of each phase must be minimal.
Differentiating~(\ref{c}) with respect to the $\theta _i $ leads
to the relation
\begin{equation}
D \tilde{\theta}_i  =
J(c_i^\theta  \tilde{\theta}_{i-1} - c_i^\theta \tilde{\theta}_{i}
-c_{i+1}^\theta  \tilde{\theta}_{i} + c_{i+1}^\theta \tilde{\theta}_{i+1}
+s_i^\theta -s_{i+1}^\theta )
\label{l}
\end{equation}
which we shall need below.

For the quadratic Hamiltonian $\tilde{\cal H}$ the energy differences
can be calculated exactly.  Let $n_{[\alpha]} = n_1$ and
$n_{[\gamma]} = n$ and label the spins within the phases $[\alpha],
[\beta]$ and $[\gamma]$ by $\alpha_i, \beta_i, \gamma_i$ respectively.
Then, using (\ref{c}) the energy of each phase relative to its value
at $D = \infty$ is
\begin{eqnarray}
n_1 E_{[\alpha]}& =&
\raisebox{-3mm}{$\stackrel{n_1}
{\stackrel{{\displaystyle\sum}}{{\scriptstyle i = 1}}}$}
\{ J {c_i^\alpha} (\tilde{\alpha}_{i-1} - \tilde{\alpha}_i +
s_i^\alpha/c_i^\alpha)^2 /2 + D \tilde{\alpha}_i^2 /2 ~ -~
J (s_i^\alpha )^2/(2 c_i^\alpha ) \}, \nonumber \\
(n-n_1) E_{[\beta]}& = &
\raisebox{-3mm}{$\stackrel{n}{\stackrel{{\displaystyle\sum}}{
{\scriptstyle i = n_1+1}}}$}
\{ J {c_i^\beta} (\tilde{\beta}_{i-1} - \tilde{\beta}_i +
s_i^\beta/c_i^\beta)^2 /2 + D \tilde{\beta}_i^2 /2 ~ -~
J (s_i^\beta )^2/(2 c_i^\beta ) \}, \nonumber \\
n E_{[\gamma]} &  = &
\raisebox{-3mm}{$\stackrel{n}{\stackrel{{\displaystyle\sum}}{
{\scriptstyle i = 1}}}$}
\{ J {c_i^\gamma} (\tilde{\gamma}_{i-1} - \tilde{\gamma}_i +
s_i^\gamma/c_i^\gamma)^2 /2 + D \tilde{\gamma}_i^2 /2 ~ -~
J (s_i^\gamma )^2/(2 c_i^\gamma ) \}. \label{g}
\end{eqnarray}
Because $[\gamma]  \equiv [\alpha\beta]$ we can choose to label
the spins in such a way that
\begin{eqnarray}
\gamma_i^0 = \alpha_i^0, &&\ \ \ \  1 \leq i \leq n_1,\nonumber \\
\gamma_i^0 = \beta_i^0, &&\ \ \ \ n_1 + 1 \leq i \leq n. \label{i}
\end{eqnarray}
and take
\begin{eqnarray}
c_i^{\alpha}=c_i^{\gamma},\ \  \ \
s_i^{\alpha}=s_i^{\gamma},&&\ \ \ \
1 \leq i \leq n_1; \nonumber \\
c_i^{\beta}=c_i^{\gamma},\ \ \ \
s_i^{\beta}=s_i^{\gamma},&&\ \ \ \
n_1+1 \leq i \leq n.
\label{eqn40}
\end{eqnarray}
This means that we may drop the $\alpha$, $\beta$, $\gamma$
superscripts on the $\{c_i\}$  and $\{s_i\}$ and the final terms in (\ref{g})
drop out when the energy difference is
calculated
\begin{eqnarray}
\lefteqn{\Delta E = }\nonumber \\
&&\raisebox{-3mm}{$\stackrel{n_1}{\stackrel{{\displaystyle\sum}}
{{\scriptstyle i = 1}}}$}
[~J \{c_i(\tilde{\gamma}_{i-1} - \tilde{\gamma}_i
+ s_i/c_i  )^2 -
 c_i (\tilde{\alpha}_{i-1} - \tilde{\alpha}_i +
s_i/c_i)^2 \}/2
+ D(\tilde{\gamma}_i^2 - \tilde{\alpha}_i^2)/2 \large ] \nonumber\\
&+&\raisebox{-3mm}{$\stackrel{n}{\stackrel{{\displaystyle\sum}}
{{\scriptstyle i = n_1+1}}}$}~
[~J \{c_i( \tilde{\gamma}_{i-1} - \tilde{\gamma}_i
+ s_i/c_i  )^2 -
 c_i (\tilde{\beta}_{i-1} - \tilde{\beta}_i +
s_i/c_i)^2 \}/2\nonumber\\
&&\ \ \ \ \ \
\ \ \ \ \ + D(\tilde{\gamma}_i^2 - \tilde{\beta}_i^2)/2 \large ].
\label{d}
\end{eqnarray}
This expression can be simplified considerably using (\ref{l}).
Recalling  the periodicity of the ground state phases which ensures
\begin{eqnarray}
\tilde{\alpha}_i = \tilde{\alpha}_{n_1+i},&\;\;\;
\tilde{\beta}_{i} = \tilde{\beta}_{n-n_1 +i},&\;\;\;
\tilde{\gamma}_i = \tilde{\gamma}_{n+i} \ \ \ \ \ \ \ \forall \ i
\label{eqn14}
\end{eqnarray}
leads after some algebra to
\begin{equation}
\Delta E = J c_1
\{ (\tilde{\alpha}_{n_1} - \tilde{\beta}_n)(\tilde{\gamma}_1 -
\tilde{\gamma}_{n_1+1} ) - (\tilde{\alpha}_1 - \tilde{\beta}_{n_1+1})
(\tilde{\gamma}_n - \tilde{\gamma}_{n_1}) \}/2.
\label{eqn15}
\end{equation}
Note that $\Delta E$ depends only on the difference between a
small number of spins.  It is this which facilitates its calculation.
The expression \ref{eqn15} is exact for the quadratic Hamiltonian
\ref{c} but only correct to leading order for the full Hamiltonian
\ref{a}. However, this will be sufficient for the calculations
presented below.

\subsection{Labelling the Spins}
\label{2.3}
The energy differences we are trying to calculate are
independent of the labelling of the spins given that the
conditions (\ref{eqn40}) hold.  In general differences such as
$\tilde{\alpha}_{n_1} - \tilde{\beta}_n$ in the energy difference
(\ref{eqn15}) will be polynomials in
$D^{-1}$.  Low order terms will cancel when the difference in
(\ref{eqn15}) is taken, in such a way that the final result becomes
independent of the labelling.  However calculationally the
problem is simplified by a careful choice of spin labels
which allow the leading order contribution to $\Delta E$ to be
obtained directly.

We first point out that every commensurate ground state of the
Hamiltonian (1) has two points of mirror symmetry in each
period evenly spaced along the chain\cite{YTC88}. It is
possible to distinguish two cases. For states of odd period
half the symmetry
points are located on lattice sites and the other half between
lattice sites.  By symmetry the spins located on the lattice
sites corresponding to mirror symmetry points do not deviate
from their $D = \infty$ position for finite $D$
($\tilde{\theta} = 0$).
For states of even period the symmetry points
are located either all between or all on lattice sites.  For
states formed by the branching process $[\alpha] + [\beta]
\Rightarrow [\alpha \beta]$ only the former can occur.

As the branching process $[\alpha] + [\beta] \Rightarrow [\gamma]
\equiv [\alpha\beta]$ proceeds states are made up in two ways\\
(i) odd + odd $\Rightarrow$ even,\\
(ii) odd + even $\Rightarrow$ odd.\\
(A moment's reflection shows that even + even $\Rightarrow$ even
never occurs because, as the ground states are formed inductively,
no neighbouring even phases ever appear.)\\

Our aim is to choose the origin for the labelling of the spins
in such a way that the formula (\ref{eqn15}) for the energy difference is
simplified.  It is necessary to consider each of the two cases (i)
and (ii) separately.\\
\noindent (i)  \underline{odd + odd $\Rightarrow$ even}\\
We recall the notation $[\alpha] + [\beta] \Rightarrow [\gamma]$
with $[\alpha]$ and $[\gamma]$ having $n_1$ and $n$ spins
respectively in each period. A convenient choice of labelling is
\begin{equation}
\tilde{\alpha}_{n_1}=0 ; ~~~ \tilde{\beta}_n = 0.
\label{l1}
\end{equation}
It will be useful later to write
the spin deviations within a period of each phase in a way that
explicitly displays the symmetry
\begin{eqnarray}
\{ \tilde{\alpha}_i \} \equiv &&\{ \tilde{\alpha}_1, \tilde{\alpha}_2 \ldots
\tilde{\alpha}_{(n_1 - 1)/2},\tilde{\alpha}_{(n_1 + 1)/2}\;
 \ldots \tilde{\alpha}_{n_1-2}, \tilde{\alpha}_{n_1-1}, \tilde{\alpha}_{n_1}
\}\nonumber \\
= &&\{ \tilde{\alpha}_1, \tilde{\alpha}_2 \ldots
\tilde{\alpha}_{(n_1-1)/2},-\tilde{\alpha}_{(n_1-1)/2}\ldots
 - \tilde{\alpha}_2,\;- \tilde{\alpha}_1, \;\;0\} \label{l2}
\end{eqnarray}
 \begin{eqnarray}
\{ \tilde{\beta}_i \}  \equiv && \{ \tilde{\beta}_{n_1+1},
\tilde{\beta}_{n_1+2} \ldots
\tilde{\beta}_{(n+n_1-1)/2},\ \  \tilde{\beta}_{(n+n_1 + 1)/2}\
\ldots\
\ \tilde{\beta}_{n-2}, \ \ \tilde{\beta}_{n -1}, \ \ \tilde{\beta}_{n} \}
\nonumber\\
=&&  \{ \tilde{\beta}_{n_1+1}, \tilde{\beta}_{n_1+2} \ldots
\tilde{\beta}_{(n+n_1-1)/2)},
-\tilde{\beta}_{(n+n_1-1)/2},\ldots - \tilde{\beta}_{n_1+2},\;
- \tilde{\beta}_{n_1+1}, \;\;0 \}. \label{l3}
\end{eqnarray}
$\langle \alpha \rangle$ and $\langle \beta \rangle$ combine to give
an even state $\langle \gamma \rangle$ with symmetry points between
$i=(n_1-1)/2$,  $i=(n_1+1)/2$
and $i=(n+n_1-1)/2$, $i=(n+n_1+1)/2$.
 \begin{eqnarray}
\{ \tilde{\gamma}_i \}  \equiv && \{ \tilde{\gamma}_1,
\tilde{\gamma}_2 \ \ldots\
\tilde{\gamma}_{(n_1-1)/2}, \tilde{\gamma}_{(n_1 + 1)/2}\  \ldots\
\tilde{\gamma}_{n_1-1}, \tilde{\gamma}_{n_1 },
\tilde{\gamma}_{n_1+1},
\tilde{\gamma}_{n_1+2}\ldots \nonumber \\
&&\tilde{\gamma}_{(n+n_1-1)/2},
\tilde{\gamma}_{(n+n_1+1)/2} \ldots
\tilde{\gamma}_{n-1},
\tilde{\gamma}_{n}\}
\nonumber\\
= && \{ \tilde{\gamma}_1, \tilde{\gamma}_2 \ldots \tilde{\gamma}_{(n_1-1)/2},
-\tilde{\gamma}_{(n_1-1)/2},\ldots - \tilde{\gamma}_2,
-\tilde{\gamma}_1,
-\tilde{\gamma}_{n},
-\tilde{\gamma}_{n-1}\ldots \nonumber \\
&&-\tilde{\gamma}_{(n+n_1+1)/2},
\tilde{\gamma}_{(n+n_1+1)/2} \ldots
\tilde{\gamma}_{n-1},
\tilde{\gamma}_{n}\}.
\label{l4}
\end{eqnarray}
Because $\tilde{\alpha}_{n_1} - \tilde{\beta}_n = 0$, (\ref{eqn15}) simplifies
immediately to
\begin{equation}
\Delta E = - Jc_1 (\tilde{\alpha}_1 -
\tilde{\beta}_{n_1+1})(\tilde{\gamma}_n - \tilde{\gamma}_{n_1})/2.
 \label{l5}
\end{equation}
\noindent (ii)  \underline{odd + even $\Rightarrow$ odd}\\
\noindent We take $[\alpha]$ odd and $[\beta]$ even.  Choosing
$\alpha_{(n_1+1)/2} = 0$
\begin{equation}
\{\tilde{\alpha}_i \} = \{\tilde{\alpha}_1, \tilde{\alpha}_2 ~\ldots
\tilde{\alpha}_{(n_1-1)/2},0,
-\tilde{\alpha}_{(n_1-1)/2},
\ldots - \tilde{\alpha}_2,- \tilde{\alpha}_1 \}. \label{l6}
\end{equation}
A consistent choice of labelling for $[\beta]$ which results in the
correct final state is to take the mirror symmetry points to lie
between spins $i=0$ and $i=1$ and between $i=(n-n_1)/2$
and $i= (n-n_1)/2 +1$.  Hence
we may write
\begin{equation}
\{ \tilde{\beta}_i \} = \{ \tilde{\beta}_{n_1+1}, \tilde{\beta}_{n_1+2}
\ldots \tilde{\beta}_{(n+n_1)/2}, - \tilde{\beta}_{(n+n_1)/2 }
\ldots - \tilde{\beta}_{n_1+2}, - \tilde{\beta}_{n_1+1} \}. \label{l7}
\end{equation}
It is immediately apparent from (\ref{l6}) and (\ref{l7}) that
\begin{equation}
\tilde{\alpha}_{n_1} - \tilde{\beta}_n =
- (\tilde{\alpha}_1 - \tilde{\beta}_{n_1+1} ).
\label{l8}
\end{equation}
Using $[\alpha]$ and $[\beta]$ to construct $[\gamma]$ will
preserve a point of mirror symmetry at $i=(n_{1}+1)/2$.  Hence
\begin{eqnarray}
 \{\tilde{\gamma}_i\} &=& \{ \tilde{\gamma}_1, \tilde{\gamma}_2
\ldots \tilde{\gamma}_{(n_1-1)/2}, 0, - \tilde{\gamma}_{(n_1-1)/2}
\ldots - \tilde{\gamma}_2, - \tilde{\gamma}_1,- \tilde{\gamma}_{n},
-\tilde{\gamma}_{n-1} \ldots\nonumber\\ &&\ \ \ \ -\tilde{\gamma}_{(n+n_1)/2},
\tilde{\gamma}_{(n+n_1)/2}, \ldots  \tilde{\gamma}_{n-1},
 \tilde{\gamma}_{n} \}
\label{l9}
\end{eqnarray}
from which it follows that
\begin{equation}
\tilde{\gamma}_n - \tilde{\gamma}_{n_1} =
- ( \tilde{\gamma}_{n_1+1} - \tilde{\gamma}_1 ). \label{l10}
\end{equation}
Using (\ref{l8}) and (\ref{l10}) the energy difference (\ref{eqn15})
simplifies to
\begin{equation}
\Delta E = - Jc_1(\tilde{\alpha}_1 - \tilde{\beta}_{n_1+1})
(\tilde{\gamma}_n - \tilde{\gamma}_{n_1}).
\label{l11}
\end{equation}

\subsection{Recursion equations}
\label{2.4}
The next step is the calculation of the spin deviations in the
formulae (\ref{l5}) and (\ref{l11}).  To do this we
start from equation (\ref{l}) which followed from minimising the ground
state energy.  Let
\begin{equation}
\tilde{\theta}_i = \frac{\theta_i^1}{D} ~+~ \frac{\theta^2_i}{D^2}
{}~+~ \frac{\theta_i^3}{D^3} +\ldots .
\label{eqn29}
\end{equation}
Substituting into (10) and equating like powers of $D^{-1}$ gives
\begin{equation}
\theta^1_i = J (s_{i} - s_{i+1} ), \label{rr1}
\end{equation}
\begin{equation}
\theta^n_i = J (c_i \theta_{i-1}^{n-1} -
c_i \theta^{n-1}_{i} - c_{i+1} \theta^{n-1}_{i}
+ c_{i+1} \theta^{n-1}_{i+1}), \;\;\;\;\;\;\;\;
n > 1. \label{rr2}
\end{equation}
Again it is necessary to consider separately the combination of
states with different symmetries.\\
\noindent (i)  \underline{odd + odd $\rightarrow$ even}\\
\noindent Note firstly that $\alpha^1_i - \beta^1_{n_i+i}$ depends
only on the $\{s^\alpha_i\}$ and $\{s^\beta_i\}$, that is only on the
value of the spins for $D=\infty$.  Let
\begin{eqnarray}
\alpha^1_i - \beta^1_{n_1+i} &=& 0, \ \ \ \ i < n_0, \label{V1}\\
\alpha^1_i - \beta^1_{n_1+i} &=& a_0, \ \ \ i = n_0 .\label{V2}
\end{eqnarray}
Then a consequence of the spin labelling and symmetry summarised
by equations (\ref{l2}) and (\ref{l3}) is that
\begin{eqnarray}
\alpha^1_{-i} - \beta^1_{n_1-i} &=& 0, \ \ \ \ \ \ \ i <  n_0, \label{V3}\\
\alpha^1_{-i} - \beta^1_{n_1-i} &=& -a_0, \ \ \  i =  n_0.\label{V4}
\end{eqnarray}

It is apparent from the recursion equations (\ref{rr2})
that after one step of
the iteration the spin differences
$\tilde{\alpha}_i-\tilde{\beta}_{n_1+i}$ and
$\tilde{\alpha}_{-i}-\tilde{\beta}_{n_1-i}$
with $i=(n_0 -1)$ will attain a non-zero
value; after a second step these differences with $i=n_0-2$ will
become non-zero and so on.
Therefore iterating $(n_0 -1)$ times gives
the leading order result
\begin{equation}
\tilde{\alpha}_1 - \tilde{\beta}_{n_1+1} = J^{n_{0}-1}
\cos (\pi/p)^{n_0-1}a_0/D^{n_0}.
\label{V5}
\end{equation}
Note that
we have assumed that $c_i = \cos(\pi/p)\;
\forall \; i$.  This is true to leading order near $\Delta = \pi/p$.

It is also appropriate to mention here that further iteration of
the linear equations will generate corrections to (\ref{V5})
which are correct for the quadratic Hamiltonian (\ref{c}) but not for the
full chiral clock model (\ref{a}).  However, these terms are not
necessary for our argument.

As a consequence of the conditions (\ref{eqn40}) and because
$[\gamma] = [\alpha\beta]$ it must also hold that
\begin{eqnarray}
\gamma^1_{i} - \gamma^1_{n_1+i} &=& 0, \ \ \ \  -n_0 < i <  n_0, \\
\gamma^1_{i} - \gamma^1_{n_1+i} &=&  a_0, \ \ \ \   i =  \pm n_0.
\end{eqnarray}
Iterating $n_0$ times using (\ref{rr2}) gives
\begin{equation}
\tilde{\gamma}_n - \tilde{\gamma}_{n_1} = 2 J^{n_{0}}
\cos (\pi/p)^{n_0}a_0/D^{n_0+1}
\label{V8}
\end{equation}
where the factor 2 appears because terms which iterate both
from the right and left along the chain contribute to this order.
Hence the energy difference (\ref{l5}) is
\begin{equation}
\Delta E = - J^{2n_0}
\cos (\pi/p)^{2n_0} a^2_0/D^{2n_0+1} .
\label{V9}
\end{equation}
(ii)  \underline{odd + even $\Rightarrow$ odd}\\
\noindent We assume as before that (\ref{V1}) and (\ref{V2})
hold.  It follows immediately from considering
(\ref{l6}) and (\ref{l7}) that
\begin{eqnarray}
\alpha^1_{i} - \beta^1_{n_1+i} = 0, &&\ \ \ \ -( n_0-1 )<i<n_0,
\label{pp}\\
\alpha^1_{n_0} - \beta^1_{n_1+n_0} =  a_0 ,&&\ \ \ \
\alpha^1_{-(n_0-1)} - \beta^1_{n_1-(n_0-1)} = - a_0 .
\end{eqnarray}
Similarly
\begin{eqnarray}
\gamma^1_i - \gamma^1_{n_1+i} &=& 0 ,\ \ \ \  -(n_0-1) < i < n_0,\\
\gamma^1_i - \gamma^1_{n_1+i} &=&   a_0 ,\ \ \ \ i = n_0, -(n_0-1).
\end{eqnarray}
Iterating $(n_0-1)$ times
\begin{equation}
\tilde{\alpha}_1 - \tilde{\beta}_{n_1+1} = J^{n_{0}-1}
\cos (\pi/p)^{n_0-1}a_0/D^{n_0}
\label{V15}
\end{equation}
from terms iterating from the right and
\begin{equation}
\tilde{\gamma}_n - \tilde{\gamma}_{n_1} =  J^{n_{0}-1}
\cos (\pi/p)^{n_0-1}a_0/D^{n_0}
\label{V16}
\end{equation}
from terms iterating from the left.
Using (\ref{l11}) the energy difference is
\begin{equation}
\Delta E =-  J^{2n_0-1}
\cos (\pi/p)^{2n_0-1} a^2_0/D^{2n_0}.
\label{V17}
\end{equation}
Both of the energy differences (\ref{V9}) and (\ref{V17}) are
negative. Hence for $p \geq 3$
all phases appear for large $D$ near the multiphase
point in agreement with the conclusions of Chou and Griffiths\cite{CG86}.

Noting that $s_i=\pm\sin(\pi/p)\ \ \forall \ i$ it is apparent by inspection
that $a_0^2 = 4J^2 \sin^2 (\pi/p)$.
We prove in Appendix A that $n_0 = [(n_{[\gamma]}+1)/2]-1$ where $ [n]$
is the integer part of $n$.
Therefore for all combinations of phases
\begin{equation}
\Delta E =-  J^{n_{[\gamma]}-2}
\cos (\pi/p)^{n_{[\gamma]}-2} a^2_0/D^{n_{[\gamma]}-1}.
\label{V18}
\end{equation}
Correction terms ${\cal O} (1/D^{n_{[\gamma]}})$ will arise from contributions
to the energy difference (\ref{eqn15}) and the recursion equations (\ref{l})
from non-harmonic terms in the Hamiltonian (\ref{a}). There will also be
correction terms within the harmonic approximation (\ref{c}) itself
arising from
further iteration of the recursion equations (\ref{l}).

\subsection{Phase Widths}
\label{2.5}

An advantage of the formalism presented above is that it allows a
calculation of the widths of the long-period phases. We define
$\Delta_{\alpha \gamma}$,
$\Delta_{\beta \gamma}$, and
$\Delta_{\alpha \beta}$ by
\begin{eqnarray}
&&E_{[\alpha]}(\Delta_{\alpha \gamma})=E_{[\gamma]}(\Delta_{\alpha \gamma}),
\label{pw1}\\
&&E_{[\beta]}(\Delta_{\beta \gamma})=E_{[\gamma]}(\Delta_{\beta \gamma}),
\label{pw2}\\
&&E_{[\alpha]}(\Delta_{\alpha \beta})=E_{[\beta]}(\Delta_{\alpha \beta}).
\label{pw3}
\end{eqnarray}
For a stable phase $[\gamma]$ $\Delta_{\alpha \beta}$ will lie between
$\Delta_{\alpha \gamma}$ and $\Delta_{\beta \gamma}$
and if the phase $[\gamma]$ is not too wide a Taylor
expansion gives
\begin{eqnarray}
&&E_{[\alpha]}(\Delta_{\alpha \gamma})=E_{[\alpha]}(\Delta_{\alpha \beta})
+E_{[\alpha]}^{\prime}(\Delta_{\alpha \gamma}-\Delta_{\alpha \beta}),
\label{pw4}\\
&&E_{[\gamma]}(\Delta_{\alpha \gamma})=E_{[\gamma]}(\Delta_{\alpha \beta})
+E_{[\gamma]}^{\prime}(\Delta_{\alpha \gamma}-\Delta_{\alpha \beta})
\label{pw5}
\end{eqnarray}
where $^{\prime}$ denotes a derivative with repect to $\Delta$.

Subtracting (\ref{pw5}) from (\ref{pw4}) and using (\ref{pw1}) gives
\begin{eqnarray}
(E_{[\alpha]}^{\prime} - E_{[\gamma]}^{\prime})
(\Delta_{\alpha \gamma}-\Delta_{\alpha \beta})
&=&
E_{[\gamma]}(\Delta_{\alpha \beta} )- E_{[\alpha]}(\Delta_{\alpha \beta})
\label{pw6}\\
&&=E_{[\gamma]}(\Delta_{\alpha \beta}) -
\frac{n_{[\alpha]}}{n_{[\gamma]}} E_{[\alpha]}(\Delta_{\alpha \beta})
-\frac{n_{[\beta]}}{n_{[\gamma]}} E_{[\beta]}(\Delta_{\alpha \beta})
\label{pw7}\\
&&\equiv\frac{\Delta E} {n_{[\gamma]}}
\label{pw8}
\end{eqnarray}
where in the penultimate step we have used (\ref{pw3}) and in the final step
the
definition of $\Delta E$, equation (\ref{eqn6}).
Writing down a similar expression for
$\Delta_{\beta \gamma}-\Delta_{\alpha \beta}$
and combining it with (\ref{pw8}) gives an expression for the width of the
phase $[\gamma]$
\begin{equation}
W_{[\gamma]} \equiv (\Delta_{\beta \gamma}-\Delta_{\alpha \gamma})
=\Delta E\{(E_{[\beta]}^{\prime} - E_{[\gamma]}^{\prime})^{-1}
-(E_{[\alpha]}^{\prime} - E_{[\gamma]}^{\prime})^{-1}\}/n_{[\gamma]}.
\label{pw9}
\end{equation}

For $D=\infty$ the energy per spin of a ground state phase at the
multiphase point $[\alpha]$ say is
\begin{equation}
E_{[\alpha]} =-J\{ l_{[\alpha]} \cos \Delta +(1-l_{[\alpha]})\cos(-2 \pi /p
+\Delta)\}
\label{pw10}
\end{equation}
where $l_{[\alpha]}$ is the fraction of nearest-neighbour ferromagnetic
bonds.
Differentiating the expression (\ref{pw10}) and similar formulae for
$[\beta]$ and $[\gamma]$ and substituting into (\ref{pw9}) gives
to leading order
\begin{equation}
W_{[\gamma]}=\frac{\Delta E (l_{[\alpha]}-l_{[\beta]})}
{2Jn_{[\gamma]}\sin(\pi/p)(l_{[\beta]}-l_{[\gamma]})
(l_{[\alpha]}-l_{[\gamma]})}.
\end{equation}
It is not hard to show inductively that
\begin{eqnarray}
l_{[\beta]}-l_{[\alpha]}=(n_{[\alpha]}n_{[\beta]})^{-1},\ \ \ &
l_{[\beta]}-l_{[\gamma]}=(n_{[\beta]}n_{[\gamma]})^{-1},\ \ \ &
l_{[\alpha]}-l_{[\gamma]}=(n_{[\alpha]}n_{[\gamma]})^{-1}.
\end{eqnarray}
Therefore using (\ref{V18})
we finally obtain
\begin{equation}
W_{[\gamma]}=-2 J^{n_{[\gamma]}-1}\cos(\pi/p)^{n_{[\gamma]}-2}
\sin(\pi/p) n_{[\gamma]}/D^{n_{[\gamma]}-1}.
\label{pw11}
\end{equation}

\section{XY model with competing first-and second-neighbour
interactions and $p$-fold spin anisotropy}
\label{3.0}
\subsection{Definitions and Notation}
\label{3.1}
In the second part of the paper we extend the formalism developed earlier
to obtain new results for a more complex spin model.  This is the
XY model with competing first- and second-neighbours interactions
and $p$-fold spin anisotropy.  Each classical XY spin vector lies
in a plane perpendicular to the $z$-axis and has unit length.
The Hamiltonian can be written
\begin{equation}
{\cal H} = \sum_i \left [ -J_1 \cos (\theta_{i-1} - \theta_i)
+ J_2 \cos (\theta_{i-2} - \theta_i) -
D (\cos (p \theta_i) -1)/p^2 \right ] \label{z1}
\end{equation}
where $\theta_i$ is the angle between the spin located at site $i$ and,
say, the $x$-axis.  Competition is introduced along the $z$-direction
by taking the first and second neighbour interactions to be
ferromagnetic and anti-ferromagnetic, respectively
$( J_1 > 0, J_2 > 0)$.  $x = J_2/J_1$ will prove an important
variable in the description of the phase diagram.

The parameter $D > 0$ models a $p$-fold spin anisotropy in the
$(x,y)$ plane.  The ground state in the two limits $D=0$ and
$D = \infty$ is well understood.  For $D=0$ it is ferromagnetic
for $x < 1/4$.  For $x > 1/4$ it exhibits
helical order with a wave vector $\vec{q} = q \vec{\hat{z}}$ which is,
in general, incommensurate with the underlying lattice.  The
magnitude of the wave vector is determined by the exchange
energies through the relation $\cos q = (4x)^{-1}$.

For $D = \infty$, however, the spin angles $\theta_i$ are
constrained to take one of the discrete set of values $2\pi n_i/p$,
$n_i = 0 , 1, \ldots p-1$.
The Hamiltonian (\ref{z1}) then reduces to a $p$-state clock model with
competing interactions.  The ground state now has a very different
character: only a few short-period commensurate phases are stable
as $x$ is varied.  Boundaries between the different ground states can
either be simple first-order transitions with only the neighbouring
phases being stable or multiphase points at which an infinite number
of phases have the same energy.

To fully describe the ground states and the phases degenerate at
the multiphase points it is necessary to extend the notation
introduced in Section \ref{2.1}.  Note that at multiphase points such as
$x = 1$ for $p = 6$ all phases with $n_{i} - n_{i-1} = 1, 2$,
with the proviso that $n_{i+1} - n_{i} = n_{i} - n_{i-1} =2$ is not
allowed, are stable.  Hence the natural definition of a wall is
as lying between spins $i$ and $i-1$ for which $n_{i} - n_{i-1} =2$.
The term band is used as before to describe the sequence of spins
between two walls.  The phases stable at the multiphase point
can then be more easily described as those containing only bands
of lengths $\geq$ 2.

A given state will be labelled by $<\ell_1, \ell_2 \ldots \ell_m >$
where the repeating sequence comprises bands of length
$\ell_1, \ell_2 \ldots \ell_m $.  It may be helpful to list some
examples for $p=6$
\begin{eqnarray}
<2>          &&        \ \     \ldots \mid 01 \mid 34 \mid 01\mid \ldots
\nonumber \\
< 2 3>           &&  \ \     \ldots \mid 01 \mid 345 \mid 12 \mid 450 \mid
\ldots
\nonumber \\
<3^24 >        &&    \ \   \ldots \mid 012 \mid 450 \mid 2345
                         \mid 123 \mid 501 \mid 3450 \mid  \ldots
\nonumber \\
<\infty>   &&   \ \      \ldots 01234 \ldots
\label{z2}
\end{eqnarray}
Compare the notation using square brackets $[ \ell_1, \ell_2 \ldots
\ell_m ]$ introduced in Section \ref{2.1}
where the walls correspond to $n_{i} - n_{i-1} = 1$ in a background
matrix of $n_{i} - n_{i-1} = 0$. This will also be needed here.

We are now in a position to describe the ground states of the
Hamiltonian (\ref{z1}) for all values of $p$ and $D=\infty$.
The results which were obtained by comparing the energies of the
possible ground states and checked using the Floria-Griffiths
algorithm\cite{FG} to ensure no states were missed
are summarised
in Figure 1.\\
(i) For $p=2$ and $3$  [$\infty$] is stable for $x < 1/2$ and [2] for $x >
1/2$.
$x = 1/2$ is a multiphase point. \\
(ii) For $p=4$ [$\infty$] is stable for $x < 1/2$ and $<\infty>\equiv [1]$
 (together with
the phase \ldots 002200 \ldots ) for $x > 1/2$.  Again $x = 1/2$
is a multiphase point. \\
(iii) For $p=5$ for $x < x^{(5)}_0 = \{1 +
\cos (2 \pi/5) \}^{-1}/2$  $[\infty]$ is stable.  For $x > x_0$, $<\infty>$ is
stable.  $x_0$ is not a multiphase point.\\
(iv) For $p \geq 6$ there is a common trend for small $x$.
For $x < {x}_1^{(p)} = \{1 + \cos (2 \pi/p)\}^{-1}/2$ [$\infty$] is
stable.  For $x_1^{(p)} < x < x_0^{(p)}
=\{\cos(2 \pi/p)-\cos(6 \pi/p)\}/2\{\cos(4 \pi/p)-\cos(6 \pi/p)\}$
the phase $<\infty>$ appears.  For $x > x^{(p)}_0$ $<2>$ is
stable.  (For $p=6$ the state \ldots 003300 \ldots is degenerate
with $<2>$.)  For $p=6$ and $7$ the phases listed above provide
all the ground states.  For $p \geq 8$ other phases are
stable for higher values of $x$, but these phases will not
concern us here.  At the points $x_1^{(p)}$ the phases $[\infty]$ and
$<\infty>$ coexist but no other phases are stable and there is
a first order transition.  $x^{(p)}_0$ however is a multiphase
point where all phases comprising bands of length $\geq$ 2
are degenerate.

Our aim is to apply an expansion in $D^{-1}$ to understand the
phase structure in the vicinity of the multiphase points $x_0^{(p)}$.
We proceed in the same way as in Section \ref{2.0}.
The results are listed in Section \ref{3.6}.
A brief account of this work has appeared elsewhere\cite{SY94}.

\subsection{The Energy Differences}
\label{3.2}
Our approach follows that described in Section \ref{2.2} for the chiral
clock model.  The first step as before is to expand the Hamiltonian
(\ref{z1}) to second order in the $\tilde{\theta}_i$, the deviations
of the spins from their positions at $D = \infty$, which were
defined by equation (\ref{eqn7}).  This gives
\begin{eqnarray}
\tilde{{\cal H}} &=& {\cal H} \mid_{D = \infty} +
\sum_i \{ J^\theta_{i,[1]} (\tilde{\theta}_{i-1} -
\tilde{\theta}_i + \Delta^\theta_{i,[1]})^2 - J_1(s^\theta_{i,[1]})^2
/c^\theta_{i, [1]} \nonumber \\
&+& J_2(s^\theta_{i,[2]})^2/c^\theta_{i,[2]} - J^\theta_{i,[2]}
(\tilde{\theta}_{i-2} - \tilde{\theta}_i
+ \Delta^\theta_{i,[2]})^2 + D \tilde{\theta}_i^2 \} /2
\label{h2}
\end{eqnarray}
where
\begin{eqnarray}
s_{i,[m]}^\theta = \sin (\theta^0_{i-[m]} - \theta^0_i ), &&\ \ \
c^\theta_{i,[m]} = \cos (\theta^0_{i-[m]} - \theta^0_i),\nonumber \\
\Delta^\theta_{i,[m]} = s^\theta_{i, [m]}/c^\theta_{i,[m]},&&\ \ \
J^\theta_{i, [m]} = J_m c^\theta_{i, [m]}.
\end{eqnarray}
Minimising (\ref{h2}) with respect to $\tilde{\theta}_i$ leads to the
equation
\begin{eqnarray}
D \tilde{\theta}_i &= &
J_{i, [1]} (\tilde{\theta}_{i-1} - \tilde{\theta}_i +
\Delta^{\theta}_{i, [1]} ) - J_{i+1, [1]}
(\tilde{\theta}_i -
\tilde{\theta}_{i+1} + \Delta^{\theta}_{i+1, [1]} )\nonumber\\
&&-J_{i, [2]} (\tilde{\theta}_{i-2} - \tilde{\theta}_i +
\Delta^{\theta}_{i, [2]}) + J_{i+2, [2]}
(\tilde{\theta}_i - \tilde{\theta}_{i+2} + \Delta^{\theta}_{i+2, [2]} ).
\label{z7}
\end{eqnarray}

It is possible to calculate the energy difference $\Delta E$,
defined by (\ref{eqn6}), exactly for the quadratic Hamiltonian~(\ref{h2}).
As before
we consider the branching process $[\alpha] + [\beta] \Rightarrow
[\alpha \beta] \equiv [\gamma]$, let $n_{[\alpha]} = n_1$
and $n_{[\gamma]} = n$ and identify the spins within the
phases $[\alpha]$, $ [\beta]$ and $[\gamma]$ by $\alpha_i,
\beta_i$ and $\gamma_i$ respectively.  We choose to label the
spins such that
\begin{eqnarray}
\gamma_i^0 &=& \alpha_i^0, \ \ \ \ \ \ \ \ 1 \leq i \leq n_1, \nonumber \\
\gamma_i^0 &=& \beta_i^0, \ \ \ \ \ \ \ \  n_i + 1 \leq i \leq n.
\end{eqnarray}
from which it follows that
\begin{eqnarray}
\Delta^\alpha_{i, [m]} = \Delta^\gamma_{i, [m]},& \ \ \
J^\alpha_{i, [m]} = J^\gamma_{i, [m]}, &\ \ \
1\le i \le n_1, \nonumber\\
\Delta^\beta_{i, [m]} = \Delta^\gamma_{i, [m]},& \ \ \
J^\beta_{i, [m]} = J^\gamma_{i, [m]},&\ \ \
n_1+1 \le i \le n.
\label{s2}
\end{eqnarray}

The energy difference (\ref{eqn6}) may now be written
\begin{eqnarray}
2 \Delta E &=& \raisebox{-3mm}{$\stackrel{n_1}
{\stackrel{{\displaystyle\sum}}{{\scriptstyle i =
1}}}$} \{ [J_{i, [1]}(\tilde{\gamma}_{i-1} - \tilde{\gamma}_i
+ \tilde{\Delta}_{i, [1]})^2 - J_{i, [1]} (\tilde{\alpha}_{i-1}
- \tilde{\alpha}_i + \Delta_{i, [2]})^2]\nonumber \\
&-& [J_{i,[2]}(\tilde{\gamma}_{i-2} - \tilde{\gamma}_i
+ \Delta_{i, [2]})^2 - J_{i, [2]} (\tilde{\alpha}_{i-2} -
 \tilde{\alpha}_i + \Delta_{i, [2]})^2 ]\nonumber \\
&+& D (\tilde{\gamma}^2_i - \tilde{\alpha}^2_i) \} \nonumber \\
&+& \raisebox{-3mm}{$\stackrel{n}{\stackrel{{\displaystyle\sum}}
{{\scriptstyle n_i + 1}}}$}
\{[J_{i, [1]}(\tilde{\gamma}_{i-1} - \tilde{\gamma}_i
+ \Delta_{i, [1]} )^2 - J_{i, [1]}(\tilde{\beta}_{i-1}
- \tilde{\beta}_i + \Delta_{i,[1]})^2] \nonumber \\
&-& [J_{i,[2]}(\tilde{\gamma}_{i-2} - \tilde{\gamma}_i
+ \Delta_{i,[2]})^2 - J_{i,[2]}(\tilde{\beta}_{i-2}
- \tilde{\beta}_i + \Delta_{i,[2]})^2 ]\nonumber \\
&+& D (\tilde{\gamma}^2_i - \tilde{\beta}^2_i)\}.
\end{eqnarray}
Using equation (\ref{z7}) this can be simplified to give
\begin{eqnarray}
2\Delta E &=& J_{1, [1]} \{ ( \tilde{\alpha}_{n_1} - \tilde{\beta}_n )
(\tilde{\gamma}_1 - \tilde{\gamma}_{n_1 + 1} ) -
(\tilde{\alpha}_1 - \tilde{\beta}_{n_1+1})(\tilde{\gamma}_n - \tilde{\gamma}
_{n_1}) \} \nonumber \\
&-& J_{1, [2]} \{ ( \tilde{\alpha}_{n_1-1} - \tilde{\beta}_{n-1} )
( \tilde{\gamma}_1 - \tilde{\gamma}_{n_1+1}) -
( \tilde{\alpha}_1 - \tilde{\beta}_{n_1+1})
( \tilde{\gamma}_{n-1} - \tilde{\gamma}_{n_1-1} ) \} \nonumber \\
&-& J_{2, [2]} \{ ( \tilde{\alpha}_{n_1} - \tilde{\beta}_n)
(\tilde{\gamma}_2 - \tilde{\gamma}_{n_1+2}) -
( \tilde{\alpha}_2 - \tilde{\beta}_{n_1+2})
(\tilde{\gamma}_n - \tilde{\gamma}_{n_1} ) \}.
\label{k1}
\end{eqnarray}
Again the energy difference depends only on the difference of the
spin deviations on a few sites.  Given a careful labelling of the
states these can be calculated to leading order and their signs determined.

\subsection{Labelling the Spins}
\label{3.3}
We label the spins in a way identical to that described in
Section \ref{2.3}.  This simplifies the formula for the energy
difference $\Delta E$ as follows:\\
\noindent (i) \underline{odd + odd $\Rightarrow$ even}\\
With the choice (\ref{l1}) it can be read off immediately
from (\ref{l2}), (\ref{l3}) and (\ref{l4}) that
\begin{eqnarray}
&&\tilde{\alpha}_{n_1} - \tilde{\beta}_n = 0,\nonumber \\
&&\tilde{\alpha}_1 - \tilde{\beta}_{n_1+1} = -(\tilde{\alpha}_{n_1-1} -
\tilde{\beta}_{n-1}),\nonumber \\
&&(\tilde{\gamma}_1 - \tilde{\gamma}_{n_1+1}) = (\tilde{\gamma}_{n} -
\tilde{\gamma}_{n_1})
\end{eqnarray}
and hence from (\ref{k1})
\begin{eqnarray}
2 \Delta E = &-& J_{1, [1]}
(\tilde{\alpha}_1 - \tilde{\beta}_{n_1+1})
(\tilde{\gamma}_n - \tilde{\gamma}_{n_1})\nonumber \\
&+&  J_{1, [2]} (\tilde{\alpha}_1 - \tilde{\beta}_{n_1+1} )
( \tilde{\gamma}_{n-1} - \tilde{\gamma}_{n_1-1}
+\tilde{\gamma}_{1}-\tilde{\gamma}_{n_1+1})
\nonumber \\
&+& J_{2,[2]} (\tilde{\alpha}_2 - \tilde{\beta}_{n_1+2})
(\tilde{\gamma}_n - \tilde{\gamma}_{n_1}).\label{k2}
\end{eqnarray}
\noindent (ii) \underline{odd + even $\Rightarrow$ odd}\\
Using (\ref{l6}), (\ref{l7}), and  (\ref{l9})
\begin{eqnarray}
\tilde{\alpha}_{n_1-1} - \tilde{\beta}_{n-1}
&=& -(\tilde{\alpha}_2 - \tilde{\beta}_{n_1+2}),\nonumber \\
\tilde{\alpha}_{n_1} - \tilde{\beta}_{n}
&=& -(\tilde{\alpha}_1 - \tilde{\beta}_{n_1+1}),\nonumber \\
(\tilde{\gamma}_{n-1} - \tilde{\gamma}_{n_1-1})
&=& - ( \tilde{\gamma}_{n_1+2} - \tilde{\gamma}_2),\nonumber \\
(\tilde{\gamma}_{n} - \tilde{\gamma}_{n_1})
&=& - ( \tilde{\gamma}_{n_1+1} - \tilde{\gamma}_1).
\end{eqnarray}
Therefore
\begin{eqnarray}
2 \Delta E = && 2 J_{1,[1]} (\tilde{\alpha}_1 - \tilde{\beta}_{n_1+1})
(\tilde{\gamma}_1 - \tilde{\gamma}_{n_1+1})\nonumber \\
&+& J_{1, [2]} \{ (\tilde{\alpha}_2 - \tilde{\beta}_{n_1+2})
(\tilde{\gamma}_1 - \tilde{\gamma}_{n_1+1}) +
(\tilde{\alpha}_1 - \tilde{\beta}_{n_1+1})
(\tilde{\gamma}_2 - \tilde{\gamma}_{n_1+2}) \}\nonumber \\
&+& J_{2, [2]} \{ (\tilde{\alpha}_1 - \tilde{\beta}_{n_1+1})
(\tilde{\gamma}_2 - \tilde{\gamma}_{n_1+2}) +
(\tilde{\alpha}_2 - \tilde{\beta}_{n_1+2})
(\tilde{\gamma}_1 - \tilde{\gamma}_{n_1+1})\}.\nonumber\\
\label{k3}
\end{eqnarray}

\subsection{Recursion equations}
\label{3.4}
We follow Section \ref{2.4} in using recursion equations to derive the leading
order terms in the energy differences (\ref{k2}) and (\ref{k3}).
Substituting (\ref{eqn29}) into the energy minimisation equations
(\ref{z7}) gives
\begin{eqnarray}
\theta^1_i  &=& J_1 (s_{i,[1]} - s_{i+1,[1]}) -
 J_2 ( s_{i, [2]} - s_{i+2, [2]} )
\label{k10} \\
\theta^n_i &=& J_{i, [1]} ( \theta^{n-1}_{i-1} - \theta^{n-1}_i)
- J_{i+1, [1]} ( \theta^{n-1}_i - \theta^{n-1}_{i+1})\nonumber \\
&-&J_{i, [2]} (\theta^{n-1}_{i-2} - \theta^{n-1}_i)
+ J_{i+2, [2]} (\theta^{n-1}_i - \theta^{n-1}_{i+2}).
\label{k11}
\end{eqnarray}
For the model with second-neighbour interactions it is necessary to
consider four different cases when calculating the energy
differences.\\
\noindent (i) \underline{odd + odd $\Rightarrow$ even; even starting
position}\\
\noindent Let
\begin{eqnarray}
 \alpha^1_i - \beta^1_{n_1+i} = 0,&&\ \ \   -2n_0 < i < 2n_0,\nonumber\\
 \alpha^1_{2n_0} - \beta^1_{n_1+2n_0} = a_0, &&\ \ \
 \alpha^1_{-2n_0} - \beta^1_{n_1-2n_0} =- a_0,
\end{eqnarray}
that is $i$ is even when $\alpha^1_i - \beta^1_{n_1+i}$ is first
non-zero. We consider $n_0 \ge 1$ throughout.
This implies
\begin{eqnarray}
 \gamma^1_i - \gamma^1_{n_1+i} = 0,&&\ \   -2n_0 < i < 2n_0,\nonumber\\
 \gamma^1_i - \gamma^1_{n_1+i} =  a_0, &&\ \   i = \pm 2n_0.
\end{eqnarray}

The recursion equations (\ref{k11}) show that
the spin deviations
$\tilde{\alpha}_i - \tilde{\beta}_{n_1+i}$ for
$i=\pm (2n_0-1)$ and $i= \pm
(2n_0 - 2)$ will be ${\cal O}(1/D^2)$; for $i=\pm(2n_0-3)$ and $i=
 \pm (2n_0-4)$ will be
${\cal O}(1/D^3)$ and so forth.  Hence
\begin{eqnarray}
\tilde{\alpha}_1 - \tilde{\beta}_{n_1+1} \sim {\cal O} (1/D^{n_0+1}), &&
\ \ \ \tilde{\gamma}_n - \tilde{\gamma}_{n_1} \sim {\cal O} (1/D^{n_0+1}),
\nonumber\\
\tilde{\gamma}_1 - \tilde{\gamma}_{n_1+1} \sim {\cal O} (1/D^{n_0+1}),
&&\ \ \
\tilde{\alpha}_2 - \tilde{\beta}_{n_1+2} \sim {\cal O} (1/D^{n_0})
\label{k12}
\end{eqnarray}
and the leading order term in the energy difference (\ref{k2}) is
\begin{equation}
2 \Delta E = J_{2, [2]} (\tilde{\alpha}_2 -
\tilde{\beta}_{n_1+2})(\tilde{\gamma}_n - \tilde{\gamma}_{n_1}).
\label{s1}
\end{equation}
{}From the recursion equations (\ref{k11}) iterating $(n_0-1)$ times
\begin{equation}
\tilde{\alpha}_2 - \tilde{\beta}_{n_1+2} =\{ a_0 (-J_2)^{n_0-1}
c^{n_w[2n_{0},2]}_{2c} c^{n_{0}-1-n_w[2n_{0},2]}_{2i}\}/D^{n_0} \label{f1}
\end{equation}
where $n_w(i,j)$ is the number of walls between $i$ and $j$ and
\begin{equation}
\cos (\alpha^0_{i-2} - \alpha^0_i) \equiv c_{2c} = \cos (6 \pi/p)
\end{equation}
if the iteration step jumps across a wall and
\begin{equation}
\cos(\alpha^0_{i-2} - \alpha^0_i) \equiv c_{2i} = \cos (4 \pi/p)
\end{equation}
if it does not.

Similarly iterating $n_0$ times and collecting equal terms from
the left and right
\begin{equation}
\tilde{\gamma}_n -\tilde{ \gamma}_{n_1} =
\{2a_0 (-J_2)^{n_0}
c_{2c}^{n_w[2n_{0},0]}c_{2i}^{n_{0}-n_w[2n_{0},0]}\}/D^{n_0+1}. \label{f2}
\end{equation}
Substituting (\ref{f1}) and (\ref{f2}) into (\ref{s1}) gives
\begin{equation}
 \Delta E =\{  J_{2,[2]}
a_0^2 (-J_2)^{2n_0-1}
c_{2c}^{n_w[2n_{0},2]+n_w[2n_{0},0]}c_{2i}^{2n_0-n_w[2n_{0},2]-n_w[2n_{0},0]-1}\}
/D^{2n_0+1}. \label{f3}
\end{equation}
$\tilde{\alpha}_n= 0$.  Therefore the band containing $i = n$ of
the state $<\alpha>$ must be of odd length.  1-bands are
forbidden.  If the band is of length 3, $J_{2, [2]} = J_2 c_{2c}$ and
$n_w[2n_{0},0] = n_w[2n_{0},2] + 1$.  If it is of length $\ge 5$, $J_{2, [2]}
= J_2c_{2i}$ and $n_w[2n_{0},0] = n_w[2n_{0},2]$.  In both cases
\begin{equation}
 \Delta E_0 =
-\{ a_0^2 (J_2)^{2n_0}
c_{2c}^{2n_w[2n_{0},0]}c_{2i}^{2n_0-2n_w[2n_{0},0]}\}
/D^{2n_0+1} \label{f4}
\end{equation}
where the relevance of the subscript on $\Delta E$ will become
apparent later.\\
\noindent(ii) \underline{odd + odd $\Rightarrow$ even; odd starting
position}\\
\noindent Let
\begin{eqnarray}
 \alpha^1_i - \beta^1_{n_1+i} = 0,&&\ \ \   -(2n_0-1) < i <
(2n_0-1),\nonumber\\
 \alpha^1_{2n_0-1} - \beta^1_{n_1+2n_0-1} = a_0, &&\ \ \
 \alpha^1_{-(2n_0-1)} - \beta^1_{n_1-(2n_0-1)} = -a_0
\end{eqnarray}
where we now consider $i$ odd for $\alpha^1_i - \beta^1_{n_1+i}$ first
non-zero. This implies
\begin{eqnarray}
 \gamma^1_i - \gamma^1_{n_1+i} = 0,&&\ \   -(2n_0-1) < i < (2n_0-1),\nonumber\\
 \gamma^1_i - \gamma^1_{n_1+i} =  a_0, &&\ \   i = \pm (2n_0-1).
\end{eqnarray}
Iterating the recursion equations (\ref{k11}) shows that
\begin{eqnarray}
\tilde{\alpha}_1 - \tilde{\beta}_{n_1+1} \sim {\cal O} (1/D^{n_0}), &&\ \ \
\tilde{\gamma}_n - \tilde{\gamma}_{n_1} \sim {\cal O} (1/D^{n_0+1}),
\nonumber\\
\tilde{\gamma}_1 - \tilde{\gamma}_{n_1+1} \sim {\cal O} (1/D^{n_0}),
&&\ \ \
\tilde{\alpha}_2 - \tilde{\beta}_{n_1+2} \sim {\cal O} (1/D^{n_0})
\end{eqnarray}
and hence the leading order term in the energy difference (\ref{k2})
is
\begin{equation}
2 \Delta E =  J_{1, [2]} (\tilde{\alpha}_1 -
\tilde{\beta}_{n_1+1})(\tilde{\gamma}_1 - \tilde{\gamma}_{n_1+1}
+\tilde{\gamma}_{n-1} - \tilde{\gamma}_{n_1-1}).
\end{equation}
{}From the recursion equations (\ref{k11}) iterating $(n_0-1)$ times
\begin{eqnarray}
\lefteqn{\tilde{\alpha}_1 - \tilde{\beta}_{n_1+1} =
\tilde{\gamma}_1 -\tilde{\gamma}_{n_1+1} =
\tilde{\gamma}_{n-1} -\tilde{\gamma}_{n_1-1} =} \nonumber \\
&&\ \ \ \ \{a_0
(-J_2)^{n_0-1}c_{2c}^{n_w[2n_{0}-1, 1]}c_{2i}^{n_0-1-n_w[2n_{0}-1, 1]}
\}/D^{n_0}.
\end{eqnarray}
$J_{1, [2]} = J_2 c_{2i}$ as the band containing the spin $i=n$ must be of
length $\geq 3$.  Therefore
\begin{equation}
 \Delta E_2 = \{ a_0^2 (J_2)^{2n_0-1}c_{2c}^{2n_w[2n_{0}-1, 1]}
c_{2i}^{2n_0-1-2n_w[2n_{0}-1, 1]}\}/D^{2n_0}.
\label{f8}
\end{equation}
\noindent(iii) \underline{odd + even $\Rightarrow$ odd; even starting
position}\\
\noindent Let
\begin{eqnarray}
\alpha_i^1 - \beta_{n_1+i}^1 = 0, &&\ \ \  -(2n_0-1) < i < 2n_0, \nonumber\\
\alpha_{2n_0}^1 - \beta_{n_1+2n_0}^1 =  a_0, &&\ \ \
\alpha_{-(2n_0-1)}^1 - \beta_{n_1-(2n_0-1)}^1 = - a_0,
\nonumber\\
\alpha_{2n_0+1}^1 - \beta_{n_1+2n_0+1}^1 =  a_1, && \ \ \
\alpha_{-2n_0}^1 - \beta_{n_1-2n_0}^1 = - a_1
\end{eqnarray}
and similarly for the $\gamma_i^1-\gamma_{n_1+i}^1$.
Noting from the recursion equations (\ref{k11}) that
\begin{eqnarray}
\tilde{\alpha}_1 - \tilde{\beta}_{n_1+1} \sim {\cal O} (1/D^{n_0+1}), &&\ \ \
\tilde{\gamma}_1 - \tilde{\gamma}_{n_1+1} \sim {\cal O} (1/D^{n_0+1}),
\nonumber\\
\tilde{\gamma}_2 - \tilde{\gamma}_{n_1+2} \sim {\cal O} (1/D^{n_0}),
&&\ \ \
\tilde{\alpha}_2 - \tilde{\beta}_{n_1+2} \sim {\cal O} (1/D^{n_0})
\end{eqnarray}
and that whatever the arrangement of walls symmetry implies $J_{1, [2]}
= J_{2, [2]}$ the energy difference (\ref{k3}) becomes
\begin{equation}
 \Delta E = J_{1, [2]} \{ (\tilde{\alpha}_2 - \tilde{\beta}_{n_1+2})
(\tilde{\gamma}_1 - \tilde{\gamma}_{n_1+1}) +  (\tilde{\alpha}_1 -
\tilde{\beta}_{n_1+1}) (\tilde{\gamma}_2 - \tilde{\gamma}_{n_1+2}) \}.
\end{equation}
Iterating the recursion equations (\ref{k11}) $(n_0-1)$ times
\begin{equation}
\tilde{\alpha}_2 - \tilde{\beta}_{n_1+2} = \tilde{\gamma}_2 -
\tilde{\gamma}_{n_1+2} = \{a_0
(-J_2)^{n_0-1}c_{2c}^{n_w[2n_{0},2]}c_{2i}^{n_0-1-n_w[2n_{0},2]}\}/D^{n_0}.
\end{equation}
Calculation of $\tilde{\alpha}_1 - \tilde{\beta}_{n_1+1}$ is
slightly more
involved as iterating $n_0$ times contributions can arise
either from hops
from $i=2n_0$ which include a $J_1$ term or from hops from $i=2n_0+1$
which include only $J_2$ terms
\begin{eqnarray}
\tilde{\alpha}_1-\tilde{\beta}_{n_1+1}&=& \{a_0
(-J_2)^{n_0-1}
J_1 c_{2c}^{n_w[2n_{0},1]-1}   c_{2i}^{n_{0}-n_w[2n_{0},1]-1}
(c_{2c} c_{1i} (n_{0}-\tilde{n}_{0}) + c_{2i} c_{1c} \tilde{n}_{0}) \nonumber\\
&+&a_1 (-J_2)^{n_0} c_{2c}^{n_w[2n_{0}+1,1]}
c_{2i}^{n_{0}-n_w[2n_{0}+1,1]}   \nonumber \\
&-& a_0
(-J_2)^{n_0} c_{2c}^{n_w[-(2n_{0}-1), 1]}
c_{2i}^{n_{0}-n_w[-(2n_{0}-1),1]}\}/D^{n_0+1}
\end{eqnarray}
where $\tilde{n}_{0}$ is the number of paths in which the $J_1$ hop crosses a
wall between $2n_{0}$ and 1, $c_{1c}=\cos (4 \pi/p)$, and $c_{1i}=\cos(2
\pi /p)$.
The expression for $\tilde{\gamma}_1 - \tilde{\gamma}_{n_1+1}$
is similar but the last term, as it corresponds to iteration from the
left, contributes with opposite sign.  Therefore the energy difference
(\ref{k3}) is
\begin{eqnarray}
\lefteqn{\Delta E_3 =}\nonumber \\
&&\ \  \{2J_{1,[2]}   a^2_0
(J_2)^{2n_0-2} J_1 c_{2c}^{n_w[2n_{0}, 2] + n_w[2n_{0},
1]-1} c_{2i}^{2n_{0}-2 - n_w[2n_{0}, 2] - n_w[2n_{0}, 1]}
( c_{2c} c_{1i} (n_{0}-\tilde{n_{0}}) + c_{2i} c_{1c} \tilde{n_{0}} )
\nonumber\\
&&\ \ - 2J_{1, [2]} a_0 a_1   (J_2)^{2n_0-1}
c_{2c}^{n_w[2n_{0}, 2] + n_w[2n_{0}+1, 1]} c_{2i}^{2n_{0}-1-n_w[2n_{0}, 2]
- n_w[2n_{0}+1, 1]}\}/D^{2n_0+1}.\nonumber \\
\label{f12}
\end{eqnarray}
\noindent(iv) \underline{odd + even $\Rightarrow$ odd; odd starting
position}\\
\noindent Let
\begin{eqnarray}
\alpha^1_i - \beta^1_{n_1+i} = 0,  &&\ \ \ - (2n_0-2) < i < 2n_0-1,
\nonumber\\
\alpha_{2n_0-1}^1 - \beta_{n_1+2n_0-1}^1 =  a_0, &&\ \ \
\alpha_{-(2n_0-2)}^1 - \beta_{n_1-(2n_0-2)}^1 = - a_0,
\nonumber\\
\alpha_{2n_0}^1 - \beta_{n_1+2n_0}^1 =  a_1, && \ \ \
\alpha_{-(2n_0-1)}^1 - \beta_{n_1-(2n_0-1)}^1 = - a_1
\end{eqnarray}
and similarly for the $\gamma_i^1-\gamma_{n_1+i}^1$.
Iterating the equations (\ref{k11}) $(n_0-1)$ times
\begin{equation}
\tilde{\alpha}_1 -\tilde{\beta}_{n_1+1} =
\tilde{\gamma}_1 - \tilde{\gamma}_{n_1+1} = \{a_0
(-J_2)^{n_0-1} c_{2c}^{n_w[2n_{0}-1,
1]}c_{2i}^{n_{0}-1-n_w[2n_{0}-1,1]}\}/D^{n_0},\nonumber
\end{equation}
\begin{eqnarray}
\lefteqn{\tilde{\alpha}_2 -\tilde{\beta}_{n_1+2}
 = \tilde{\gamma}_2 - \tilde{\gamma}_{n_1+2} }\nonumber \\
&&= \{a_0
(-J_2)^{n_0-2}J_1c_{2c}^{n_w[2n_{0}-1, 2]-1} c_{2i}^{n_{0}-n_w[2n_{0}-1,
2]-2}
(c_{2c} c_{1i}(n_{0}-1-n_{0}^{\prime}) + c_{2i} c_{1c}n_{0}^{ \prime} )
\nonumber\\
&& \;\;\;\;\;\;+a_1 (-J_2)^{n_0-1}c_{2c}^{n_w[2n_{0},2]}
c_{2i}^{n_{0}-1-n_w[2n_{0},2]}\}/D^{n_0}
\end{eqnarray}
where $n_{0}^{\prime}$ is the number of paths in which the $J_1$ hop crosses a
wall between $2n_{0}-1$ and 2.  Using these formulae the energy difference
(\ref{k3}) is
\begin{eqnarray}
\lefteqn{\Delta E_1 =}\nonumber \\
&&\{ -J_{1, [1]} a_0^2
(J_2)^{2n_0-2}c_{2c}^{2n_w[2n_{0}-1,
1]}c_{2i}^{2n_{0}-2-2n_w[2n_{0}-1,1]}\nonumber\\
&& -2J_{1,[2]} a_0^2 (J_2)^{2n_0-3}
J_1c_{2c}^{n_w[2n_{0}-1,2]+n_w[2n_{0}-1,1]-1}
c_{2i}^{2n_{0}-3-n_w[2n_{0}-1,2]-n_w[2n_{0}-1,1]}\nonumber \\
&&\ \ \ \ \ \ \ \ \ \ \ \ \ \ \ \ \ \ \ \ \ \ \ \ \ \
(c_{2c}c_{1i} (n_{0}-1-n_{0}^{\prime}) + c_{2i}c_{1c}n_{0}^{\prime })
\nonumber\\
&&+2J_{1, [2]} a_0a_1 (J_2)^{2n_0-2}
c_{2c}^{n_w[2n_{0}-1,1]+n_w[2n_{0}, 2]}c_{2i}^{2n_{0}-2-n_w[2n_{0}-1,1]
- n_w[2n_{0}, 2]}\}/D^{2n_0}.\nonumber \\
\label{f16}
\end{eqnarray}
\subsection{Phases bounding $\langle \infty \rangle$}
\label{3.5}
The formulae for the energy differences calculated in the previous
section may no longer hold when one of the initial phases is $\langle
\infty \rangle = \ldots 0123 \ldots$. Therefore this case is now
treated separately. As the phase diagram is constructed recursively
the phases which will appear on the $\langle \infty \rangle$ boundary
are $\langle m \rangle$, $m=3,4,5 \ldots$. Hence the energy
differences we wish to calculate are
\begin{equation}
\Delta E^{\langle \infty \rangle}
=(m+1)E_{\langle m+1 \rangle}
-mE_{\langle m \rangle}-
E_{\langle \infty \rangle}.
\end{equation}
In the $\langle \infty \rangle$ phase the spins remain in their clock
positions as $D$ is reduced from $\infty$.

We must again consider four possibilities.\\
(i) \underline{$\langle 4n+3 \rangle + \langle \infty \rangle \Rightarrow
\langle 4n+4 \rangle $}\\
We consider $n \ge 1$ throughout.
Choose $\langle \beta \rangle = \langle \infty \rangle$. Then
$\tilde{\beta}=0$ and $n_1=n-1$. The initial conditions are
\begin{eqnarray}
&\alpha_i^1=0,\ \ \ \ -2n<i<2n,&\\
&\alpha_{2n}^1=-\alpha_{-2n}^1=a_0,&\\
&\gamma_i^1=0,\ \ \ \ -(2n+1)<i<2n,&\\
&\gamma_{2n}^1=-\gamma_{-(2n+1)}^1=a_0.&
\end{eqnarray}
Noting that $\gamma_{n_1}=-\gamma_{n}$ the calculation
follows that described in Section \ref{3.4}(i). The energy difference is given
by equation (\ref{f4}) with $n_w[2n,0]=0$
\begin{equation}
\Delta E_0^{\langle \infty \rangle}=
-\{a_0^2 (J_2)^{2n} c_{2i}^{2n}\}/D^{2n+1}.
\label{g4}
\end{equation}
(ii) \underline{$\langle 4n+1 \rangle + \langle \infty \rangle \Rightarrow
\langle 4n+2 \rangle$} \\
Again choose $\langle \beta \rangle = \langle \infty \rangle$.
The initial conditions are now
\begin{eqnarray}
&\alpha_i^1=0,\ \ \ \ -(2n-1)<i<2n-1,&\\
&\alpha_{2n-1}^1=-\alpha_{-(2n-1)}^1=a_0,&\\
&\gamma_i^1=0,\ \ \ \ -2n<i<2n-1,&\\
&\gamma_{2n-1}^1=-\gamma_{-2n}^1=a_0.&
\end{eqnarray}
Recalling that $n_1 = n-1$,
\begin{eqnarray}
&\tilde{\gamma}_1 \sim \tilde{\gamma}_{n_1-1} \sim \tilde{\alpha}_1
\sim \tilde{\alpha}_2 \sim {\cal O}(1/D^{n-1}),&\nonumber\\
&\tilde{\gamma}_{n-1} \sim \tilde{\gamma}_{n_1} \sim \tilde{\gamma}_n
\sim \tilde{\gamma}_{n_1+1} \sim {\cal O}(1/D^n)
\end{eqnarray}
and that the symmetry of $\langle \gamma \rangle$ implies
\begin{equation}
\tilde{\gamma}_1 = -\tilde{\gamma}_{n_1-1}
\end{equation}
the leading term in the energy difference (\ref{k2}) is
\begin{equation}
\Delta E = 2J_{1, [2]} \tilde{\alpha}_1 \tilde{\gamma}_1.
\end{equation}
Iterating the recursion equations (\ref{k11}) $(n-1)$ times to obtain
$\tilde{\alpha}_1$ and $\tilde{\gamma}_1$ gives a result
\begin{equation}
\Delta E_2^{<\infty>} = \{a^2_0 (J_2)^{2n-1} c_{2i}^{2n-1}
\}/D^{2n}.
\label{g8}
\end{equation}
\noindent (iii) \underline{$\langle 4n+2 \rangle + \langle \infty
\rangle
\Rightarrow \langle 4n+3 \rangle$}\\
To use the energy difference formula (60)
$\langle\beta\rangle$ must be even.  Therefore we choose
$\langle\alpha\rangle = \langle\infty\rangle$ implying $n_1=1$ and
$\tilde{\alpha}_i = 0 ~\forall ~i$.  The initial conditions are
\begin{eqnarray}
{\beta}_{n_1+i}^1 = 0, &&  \ \ \ -(2n-1) < i < 2n,\nonumber\\
{\beta}_{n_1+2n}^1 = - {\beta}_{n_1-(2n-1)}^1 = a_0 &&\ \ \
{\beta}_{n_1+2n+1}^1 = - {\beta}_{n_1-2n}^1 = a_1 \nonumber\\
{\gamma}_i^1 = 0, &&\ \ \  - (2n-1) < i < (2n+1),\nonumber\\
{\gamma}_{2n+1}^1 = - {\gamma}_{-(2n-1)}^1 = a_0,&&\ \ \
{\gamma}_{2n+2}^1 = -{\gamma}_{-2n}^1 = a_1.
\end{eqnarray}
Using the symmetry properties of $\langle \beta \rangle$
and $\langle \gamma \rangle$
\begin{eqnarray}
\tilde{\beta}_{n_1+i} = -\tilde{\beta}_{n_1-i+1}, &&\ \ \
\tilde{\gamma}_i = -\tilde{\gamma}_{-i+2} \label{zz}
\end{eqnarray}
and noting that
\begin{eqnarray}
\tilde{\beta}_3 \sim \tilde{\gamma}_3 \sim {\cal O}(1/D^{n-1}), &&\ \ \
\tilde{\beta}_2 \sim \tilde{\gamma}_2 \sim {\cal O}(1/D^{n})
\end{eqnarray}
the leading order contribution to the energy difference (\ref{k3}) is
\begin{equation}
2 \Delta E = (J_{1, [2]} + J_{2, [2]}) (\tilde{\beta}_3 \tilde{\gamma}_2 +
\tilde{\beta}_2 \tilde{\gamma}_3).
\end{equation}
Iterating the recursion equations (\ref{k11}) to obtain the spin
deviations the final result is
\begin{equation}
\Delta E_3^{\langle\infty \rangle} = J_2^{2n-1} c_{2i}^{2n-1}
\{a_0^2(2J_1 n c_{1i} + J_2 c_{2i} ) - 2a_0 a_1 J_2 c_{2i}
\}/D^{2n+1}.
\label{g12}
\end{equation}
\noindent (iv) \underline{$\langle 4n \rangle + \langle \infty
\rangle
\Rightarrow \langle 4n+1 \rangle$}\\
Choosing
$\langle\alpha\rangle = \langle\infty\rangle$ the initial conditions are
\begin{eqnarray}
{\beta}_{n_1+i}^1 = 0, &&  \ \ \ -(2n-2) < i < 2n-1,\nonumber\\
{\beta}_{n_1+2n-1}^1 = - {\beta}_{n_1-(2n-2)}^1 = a_0 &&\ \ \
{\beta}_{n_1+2n}^1 = - {\beta}_{n_1-(2n-1)}^1 = a_1 \nonumber\\
{\gamma}_i^1 = 0, &&\ \ \ - (2n-2) < i < 2n,\nonumber\\
{\gamma}_{2n}^1 = - {\gamma}_{-(2n-2)}^1 = a_0,&&\ \ \
{\gamma}_{2n+1}^1 = -{\gamma}_{-(2n-1)}^1 = a_1.
\end{eqnarray}
Using the symmetry properties (\ref{zz})
the leading order contribution to the energy difference (\ref{k3}) is
\begin{equation}
2 \Delta E =
2J_{1,[1]}\tilde{\beta}_2\tilde{\gamma}_2+
(J_{1, [2]} + J_{2, [2]}) (\tilde{\beta}_3 \tilde{\gamma}_2 +
\tilde{\beta}_2 \tilde{\gamma}_3-
\tilde{\beta}_2 \tilde{\gamma}_2).
\end{equation}
Using the recursion equations (\ref{k11}) we obtain
\begin{equation}
\Delta E_1^{\langle\infty \rangle} = -J_2^{2n-2} c_{2i}^{2n-2}
\{a_0^2(J_1  c_{1i}(2n-1) + J_2 c_{2i} ) - 2a_0 a_1 J_2 c_{2i}
\}/D^{2n}.
\label{g16}
\end{equation}

There will be correction terms to the formulae for the energy differences
which arise both from non-harmonic terms in the Hamiltonian (\ref{z1})
and from further iteration of the recursion equation (\ref{k11}).
These will be more dangerous than for the chiral XY model as they may
carry an additional factor $n_0^2/D$ where the $n_0^2$ comes from say
dividing a $J_2$ step into two $J_1$ steps and placing them at any
position along the chain. Therefore for large $n_0$ these terms could
dominate, possibly changing the sign of the $\Delta E$ and the sequence
of phases could terminate for any finite $D$. Analytic calculation of
the correction terms would be prohibitively difficult. However numerical
results show no sign of deviations from the leading behaviour
for $n_{[\gamma]}$=15.
A similar mechanism has been desribed for the ANNNI model at finite
temperatures\cite{SF87}.

To apply the formulae for the energy differences it is necessary to
ascertain which $\Delta E_k$ and $\Delta E_k^{\langle \infty \rangle}$,
$k=1,2,3,4$
should be used at each step of the iteration procedure. By checking
low order examples or by an inductive argument similar to that given
in Appendix A one finds that $\Delta E_k$ and $\Delta E_k^{\langle \infty
\rangle}$ are
relevant for a final phase with $n_{[\gamma]}=k\ [mod 4]$.

The values of $a_0$ and $a_1$
\begin{eqnarray}
&&a_0=J_2 \{\sin (6\pi/p)-\sin (4\pi/p)\}/D \nonumber \\
&&a_1=-J_1 \{\sin (4\pi/p)-\sin (2\pi/p)\}/D
\end{eqnarray}
follow immediately from (\ref{k11}).

\subsection{The Sequence of Phases}
\label{3.6}
The stability of the short-period phases is most easily checked by an
explicit evaluation of their energy using the Hamiltonian
(\ref{h2}). Phases which appear ${\cal O}(1/D)$ and ${\cal O}(1/D^2)$
are shown schematically in Figure 2. Stable boundaries are denoted by
a vertical line; boundaries at which an infinite number of phases
still coexist
${\cal O}(1/D^2)$ by a star. By using the energy differences derived
above it is possible to calculate which of these phases are in fact
stable when higher order terms in $1/D$ are considered. The results
depend sensitively on $p$ and so we consider each value in turn.\\
\underline{$p=6$}\\
For $p=6$ the calculation of terms ${\cal O}(1/D^2)$ establishes that
the $\langle 2 \rangle:\langle 3 \rangle$ boundary is stable. However
all other boundaries are multiphase lines and therefore any phase
which contains only bands of lengths $\ge 3$ may appear in the phase
diagram. To understand which of these phases are stable we need the
signs of the energy differences.

The energy differences are simplified by noting that for $p=6$,
$a_1=0$. To obtain the sign of $\Delta E_1$ given by equation
(\ref{f16}) two cases must be considered. If there is a wall on the axis
of symmetry (i.e. between $\alpha_0$ and $\alpha_1$)
then, recalling that all bands are of length at least 3,
$J_{1,[1]}=J_1c_{1c}$, $J_{1,[2]}=J_2 c_{2c}$, and
$n_w[2n_{0}-1,2]=n_w[2n_{0}-1,1]$. Putting in values for $c_{2c}$, etc. gives
\begin{equation}
\Delta E_1 = 3 J_1J_2^{2n_0}
(1/2)^{\{2n_{0}-2n_w-1\}}(5/4-n_{0}+3n_0^{\prime}/2)/D^{2n_0}.
\label{m1}
\end{equation}
If there is no wall on the axis of symmetry
$J_{1,[1]}=J_1c_{1i}$, $J_{1,[2]}=J_2 c_{2i}$, and
$n_w[2n_{0}-1,2]=n_w[2n_{0}-1,1]$. Therefore
\begin{equation}
\Delta E_1 = 3 J_1J_2^{2n_0}
(1/2)^{\{2n_{0}-2n_w+1\}}(1-2n_{0}+3n_0^{\prime})/D^{2n_0}.
\label{m2}
\end{equation}
which is negative.
Using similar arguments, or by inspection, all the
other energy differences are found to be negative\cite{INFBOUND}. Therefore all
phases which contain only bands of lengths $\ge 3$ appear in the phase
diagram.\\
\underline{$p=7$}\\
For $p=7$ it can be shown, using arguments similar to those presented
for $p=6$ that all the energy differences are negative for large
enough $n_0$. The cases where $n_0$ is small and $n_0^{\prime}$ and
$\tilde{n}_0$ are close in
value to $n_0$ must be checked independently. One finds that all the relevant
cases are negative except $\Delta E_1$ for $n_0=2$, $n_0^{\prime}=1$. This
determines the stability of $\langle 2^3 3\rangle$. Therfore for $p=7$
the $\langle 2 \rangle : \langle 223 \rangle$ boundary is not
split. All phases lying between $\langle 223\rangle$ and $\langle
\infty \rangle$ are however stable.\\
\underline{$p=8$}\\
The energy differences $\Delta E_0$, $\Delta E_1$,
$\Delta E_2$, and $\Delta E_3$ given by equations (\ref{f4}), (\ref{f16}),
(\ref{f8}), and (\ref{f12}) are zero. Therefore higher order terms which
are prohibitively difficult to calculate are needed to establish the
signs of the energy differences.

However, it is possible to obtain an expansion of the energies of the
low order phases numerically directly from the Hamiltonian
(\ref{z1}). (The quadratic approximation (\ref{h2}) to the Hamiltonian
may not be sufficient to pick up the correct leading behaviour in this
case.
Comparing these energies one finds that at least all phases
expected to appear ${\cal O} (1/D^5)$ $(n_{\langle \gamma
\rangle }\le 12)$ are stable.\\
\underline{$p=9$}\\
For $p=9$ the situation is complicated and no clear pattern of phases
emerges. Results ${\cal O}(1/D^2)$ indicate that only phases which are
made up of bands of lengths $\le 4$ can appear. One finds by
inspection $\Delta
E_0<0$ (equation (\ref{f4}));
$\Delta E_2>0$ (equation (\ref{f8})). For large enough $n_0$
if there is a wall on the axis of
symmetry $\Delta E_3<0$ and $\Delta E_1>0$
from equations (\ref{f12}) and (\ref{f16}) whereas if there is no wall on the
axis of symmetry $\Delta E_3>0$ and $\Delta E_1<0$. However for low
values of $n_0$ where $\tilde{n}_0$ and $n_0^{\prime}$
are close in value to $n_0$ these conclusions
may not hold and each case must be treated independently. In
particular for $\langle 23 \rangle + \langle 2 \rangle \Rightarrow
\langle 223 \rangle $ which corresponds to the case $\Delta E_3$ with
no wall on the axis of symmetry and for
$\langle 2^23 \rangle + \langle 2 \rangle \Rightarrow
\langle 2^33 \rangle $
which corresponds to the case $\Delta E_1$ with
a wall on the axis of symmetry the energy differences are negative.

Hence ${\cal O} (1/D^5)$ the phase sequence is
\begin{equation}
\langle \infty
\rangle:
\langle 4 \rangle:\langle 34 \rangle:\langle 3 \rangle;
\langle 2333 \rangle;\langle 233 \rangle;\langle 23 \rangle;
\langle 23223 \rangle;\langle 223 \rangle;\langle 2223 \rangle:
\langle 2 \rangle
\end{equation}
where : denotes a stable boundary and ; a boundary
which may be split at higher orders of the expansion.\\
\underline{$p=10$}\\
For $p=10$, $a_0=0$ which means that the leading order terms in the
energy differences are zero and the series analysis breaks down.
Numerically we have been able to show
that the only stable phases are $\langle 2^k3\rangle$ appearing
between $\langle 2 \rangle$ and $\langle 3 \rangle$. The existence of
these phases has been checked for $k \le 5$ by expanding the energies
${\cal O}(1/D^6)$.\\
\underline{$p=11$}\\
A calculation ${\cal O} (1/D)$ shows that the $\langle 2 \rangle :
\langle \infty \rangle$ boundary is stable; no new phases appear near
$x_0$ and the transition is first order.\\

\section{Discussion}
\label{4.0}
Models with competing interactions may have ground states which include
special points, so-called multiphase points\cite{FS80},
where the ground state is infinitely degenerate. At these points one might
expect that small perturbations can have a drastic effect. One possible such
perturbation is to allow the spins to soften, others are temperature\cite{SF87}
or quantum fluctuations\cite{HMY95}.

The aim of this paper has been to study the first of these possibilities.
We have considered continuous spin models with a $p$-fold spin
anisotropy $D$ which in the $D\rightarrow \infty$ limit exhibit a
multiphase point. An expansion in $1/D$ is described which
allows us to calculate the form of the phase diagram near the
multiphase points as the spins soften.

The first model we considered was the chiral XY model with $p$-fold
spin anisotropy, which for large $D$ can be thought of as a soft chiral
clock model. This model provided a useful illustation of the technique.
We showed in agreement with Chou and Griffiths\cite{CG86}
that for $p \ge 3$ all phases formed by combining adjacent structures
are stable near the multiphase point and obtained leading order
expressions for the widths of the phases.

We then described results for the more complicated situation of the XY
model with first- and second-neighbour interactions
for large $D$. Here the situation is very complex with
the behaviour near the multiphase point being strongly dependent on the
value of $p$. For $6\le p \le 10$ infinite sequences of phases are
stable, but their quantitative form is different for different $p$. For
$p \ge 11$ the phase boundary emanating from the multiphase point is
first order.

We note that not only does the symmetry of the anisotropy have a
non-universal effect on the nature of the phase diagram but also that
the physical form of the perturbation is important. For example, for the ANNNI
model itself temperature leads to a phase sequence $[ 2^k3 ]$,
$k=0,1,2 \ldots$\cite{FS80,SF87}, quantum fluctuations result in a phase
sequence $[k ]$, $k=2,3,4 \ldots$\cite{HMY95} whereas when the
spins soften there is a single first-order transition.

The models descibed here have a complex mathematical structure but also
interesting applications to real systems. For example rare-earth
magnetism has been modelled by an XY model with competing
interactions and 6-fold spin anisotropy\cite{JM91}.
Modulated structures in UNi$_2$Si$_2$ have been described
using a model expected to behave in a similar way to those considered
here\cite{MPC92}.

Finally we comment on some further possible uses for the technique introduced
in this paper. Bassler, Sasaki and Griffiths\cite{BSG92} have descibed an
upsilon point, a checkerboard structure of long-period phases which is
in some sense a two-dimensional version of the sequences of phases
we have been concerned with here.
Sasaki\cite{Sas92} found some evidence for the existence of
such a point in a spin model by performing an expansion in $1/D$ to order
$1/D^2$. An expansion taken to all orders will give firmer proof
of where such points can be found\cite{MYS95}. A similar technique can be
used to investigate interface unbinding where spin softening can allow an
interface to unbind from a surface through a series of layering
transitions\cite{MY94}.\\
{}~\\
{}~\\
\noindent{\bf Acknowledgements}: We thank Cristian Micheletti for helpful
discussions. JMY acknowledges the support of an EPSRC Advanced Fellowship
and the hospitality of the Dipartimento di Fisica, University of Padova.

\newpage

\section*{Appendix A}

We aim here to show how $n_0$ defined by equations (\ref{V1})
and (\ref{pp}) is related to $n_{[\gamma]}$ the length of the final
phase in the process $[\alpha]+[\beta] \Rightarrow [\gamma]$.

Consider two odd phases $[\alpha]$ and $[\beta]$. Let $n_0
\equiv n_0([\alpha \beta])$ for $[\alpha]+[\beta] \Rightarrow
[\alpha \beta]$.
Using the labelling scheme defined in Section \ref{2.3}(i)
\begin{equation}
[\alpha]=(\{\alpha\},-\{\alpha\},0)
\end{equation}
where we use $\{\alpha\}$
as shorthand for $\{\alpha_1,\alpha_2 \ldots \alpha_{(n_1-1)/2}\}$ (see
equation(\ref{l2})). Similarly
\begin{equation}
[\beta]=(\{\beta\},-\{\beta\},0)\label{xx}
\end{equation}
The resulting even phase is
\begin{equation}
[\alpha \beta]=(\{\alpha\},-\{\alpha\},0,\{\beta\},-\{\beta\},0).\label{yy}
\end{equation}

Consider now the process $[\alpha \beta]+[\beta] \Rightarrow
[\alpha^2 \beta]$. Now an even and odd state are combined and
therefore the labelling scheme defined in Section \ref{2.3}(ii) is
appropriate. From (\ref{xx}) and (\ref{yy})
\begin{eqnarray}
&&[\alpha \beta]=(-\{\beta\},0,\{\alpha\},-\{\alpha\},0,\{\beta\}),
\nonumber \\
&&[\beta]=(-\{\beta\},0,\{\beta\}).
\label{aa}
\end{eqnarray}
By inspection of (\ref{aa}) it is apparent that
\begin{equation}
n_0([\alpha \beta^2])=n_0([\alpha \beta])+(n_{[\beta]}+1)/2,
\label{bb}
\end{equation}
that is adding an odd state $[\beta]$ to an even state increases $n_0$
by $(n_{[\beta]}+1)/2$.

A similar argument shows that for $[\alpha]$ odd and $[\beta]$ even
\begin{eqnarray}
&&n_0([\alpha \beta^2])=n_0([\alpha \beta])+(n_{[\beta]})/2,
\label{cc}\\
&&n_0([\alpha^2 \beta])=n_0([\alpha \beta])+(n_{[\alpha]}-1)/2,
\label{dd}
\end{eqnarray}
that is adding an even (odd) state $[\beta]$ ($[\alpha]$) to an odd
state $[\alpha \beta]$
increases $n_0$ by $n_{[\beta]}/2$ ($(n_{[\alpha]}-1)/2$).

It is not hard to check that the conditions (\ref{bb}), (\ref{cc}),
and (\ref{dd}) are consistent with
\begin{eqnarray}
n_0=\{n_{[\gamma]}-(2m+1)\}/2&&\ \ \ \ n_{[\gamma]} \;\;odd \nonumber\\
n_0=\{n_{[\gamma]}-(2m+2)\}/2&&\ \ \ \ n_{[\gamma]} \;\;even
\end{eqnarray}
for any integer $m$. By inspection $n_0([23])=2$. Therefore $m=0$ and
\begin{equation}
n_0=\left[\frac{n_{[\gamma]}+1}{2}\right]-1
\end{equation}
where $[n]$ is the integer part of $n$.

\newpage
\section*{Figure Captions}
{}~\\
\noindent Figure 1: Ground states of the $p$-state clock model with
ferromagnetic first-neighbour interactions $J_1$
and antiferromagnetic second-neighbour interactions $J_2$.
A vertical line represents a first-order boundary and a star a multiphase
point.\\
{}~\\
{}~\\
\noindent Figure 2: Phase diagram of the soft clock model with competing
interactions near the multiphase point $x_0^{(p)}$ showing the phases that are
stabilised by terms in the energy ${\cal O} (1/D)$ (labelled above the line)
and ${\cal O} (1/D^2)$ (labelled below the line). A vertical line represents
a first-order boundary and a star a multiphase line at which an infinite number
of phases remain degenerate to this order.

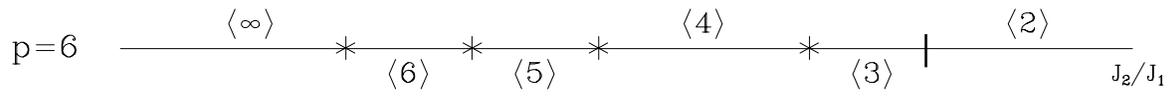
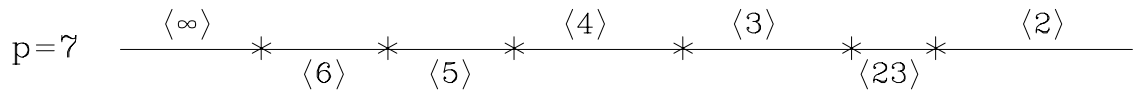
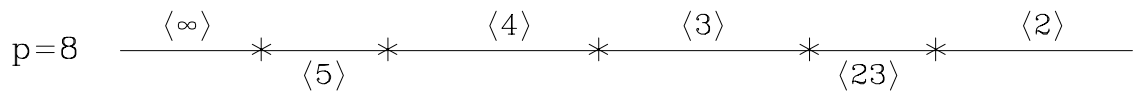
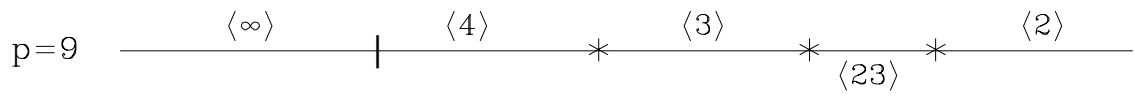
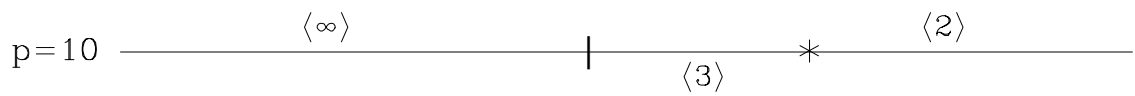
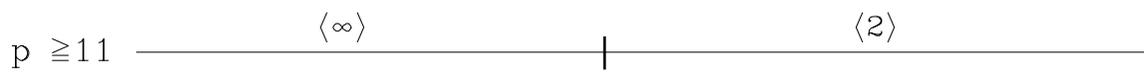
\end{document}